\nonstopmode

\documentclass[prb,twocolumn,showkeys ]{revtex4}
\usepackage{graphicx}
\usepackage{floatflt}
\usepackage{subfig}
\usepackage{amsmath,amssymb}							% mehr Mathe
\usepackage{siunitx}
\usepackage[pdftex,colorlinks=true,linkcolor=black,citecolor=black,filecolor=black,urlcolor=black]{hyperref}
\begin{document}
\title{Electronic and magnetic properties of $\text{Co}_{\text{n}}  
\text{Mo}_\text{m}$ nanoclusters with n\,+\,m\,=\,x and 2\,$\le$\,x\,$\le$\,6 atoms from DFT calculations}

\author{Simon Liebing}
 \altaffiliation{TU Bergakademie Freiberg, Institut of Theoretical Physics, 
 Leipziger Strasse 23, D-09596 Freiberg, Germany, 
 simon.liebing@physik.tu-freiberg.de}
\author{Claudia Martin}
 \altaffiliation{TU Bergakademie Freiberg, Institut of Theoretical Physics, 
 Leipziger Strasse 23, D-09596 Freiberg, Germany}
\author{Kai  Trepte}
 \altaffiliation{TU Bergakademie Freiberg, Institut of Theoretical Physics, 
 Leipziger Strasse 23, D-09596 Freiberg, Germany}
\author{Jens Kortus}
 \altaffiliation{TU Bergakademie Freiberg, Institut of Theoretical Physics, 
 Leipziger Strasse 23, D-09596 Freiberg, Germany}
\noaffiliation
\date {\today}%{17.12.2011}


\begin{abstract}
We present the results of the density functional theory study of $\text{Co}_{\text{n}}\text{Mo}_\text{m}$ nanoclusters with n+m\,=\,x and 2$\le$x$\le$6 atoms on the all-electron level using the
generalized gradient approximation. The discussion of the properties of the pure cobalt and molybdenium cluster is followed by an analysis of the respective mixed clusters of each cluster size
x. We found that the magnetic moment of a given cluster is mainly due to the Co content and increases with increasing n. The magnetic anisotropy on the other hand becomes smaller for larger magnetic moments S. We observe an increase in the binding energy, electron affinity, and average bond length with increasing cluster size as well as a decrease in the ionization potential, chemical potential, molecular hardness and the HOMO-LUMO gap. 
\end{abstract}

\maketitle

\section{Introduction}
\label{intro}
Transition metal clusters are of wide interest in various areas of research and application, for example in molecular electronics, long-time magnetic data storage or in the wide field of
catalysis \citep{Ferrando2008}. Cobalt clusters proved very interesting in terms of magnetism. There are reports of a very large magnetic anisotropy for the Co-dimer
\citep{gambardella_giant_2003,fritsch_transition_2008,strandberg_calculation_2008} making them very interesting for possible applications in future storage devices for example in combination with
hexagonal carbon rings \citep{xiao_co_2009}. It has also been shown that Mo$_2$X$_2$ (X\,=\,Fe,Co,Ni) are able act as a spin-filter \citep{fuenta2012}. Garcia-Fuente et al. \citep{fuenta2009}
computed free-standing Mo$_{4-x}$Fe$_x$ clusters and came to the conclusion that they are good candidates for molecular electronic devices. These clusters are also used widely as catalysts. In this
field the Co-Mo clusters are also known to have a strong catalytic effect, for example on the formation of carbon nanotubes \citep{Kitiyanan2000} or for hydrosulfuration
\citep{hinnemann2008,Kibsgaard2010} (using Co-Mo-S clusters). All these works show that it is imperative to understand the complex interaction of structural and electronic degrees of freedom as well
as
the influence of different chemical compositions of mixed cluster on the properties of the respective transition metal clusters. These properties govern the possible applicability of the clusters. 
There are also a few studies on mixed cobalt clusters, for example in combination with manganese \citep{ganguly2008} and copper \citep{perez_physical_2012}. Here we present a density functional theory
(DFT) study of the $\text{Co}_{\text{n}}\text{Mo}_\text{m}$ nanoclusters with n+m\,=\,x and 2$\le$x$\le$6 atoms on the all-electron level using the generalized gradient approximation (GGA). \\
There are many studies available concerning the electronic and structural properties of pure cobalt
\citep{castro_structure_1997,fan_geometry_1997,jamorski_structure_1997,pereiro2001,wang_low-lying_2005,strandberg_transition-metal_2007,datta2007,sebetci_cobalt_2008,fritsch_transition_2008,
strandberg_calculation_2008} and molybdenium (see \citep{diez2000,zhang2004,gran2008} and references therein) clusters on a theoretical and experimental level. The results of these studies will be
compared to our results for the pure cobalt and molybdenium clusters starting from dimers (x\,=\,2). This will be followed by an analysis of the respective mixed clusters of each cluster size x. In the
end we will discuss general trends in structural details as well as in the magnetic and electronic properties.

\section{Computational details}
\label{sec:1}
The starting point to construct metal clusters were the so called finite sphere packings of different sizes \citep{hoare_statistical_1976,arkus_minimal_2009,arkus_deriving_2011,hoy_structure_2012}. 
These lead to quite compact clusters. As a result configurations that are far from a spherical structure will be missed. Such structures can be relevant as known from
atoms \citep{sebetci_cobalt_2008,perez_physical_2012} and cores \citep{tohsaki_alpha_2001}. All clusters were constructed from scratch with the program package Avogadro
\citep{hanwell_avogadro:_2012} and optimized via the internal force field mechanisms. With these pre-prepared structures an unrestricted, all-electron DFT geometry optimization was
performed using the NRLMOL program package
\citep{pederson_variational_1990,jackson_accurate_1990,pederson_pseudoenergies_1991,pederson_local-density-approximation-based_1991,perdew_atoms_1992,porezag_infrared_1996,porezag_development_1997,
kortus_magnetic_2000,pederson_strategies_2000}. NRLMOL uses an optimized Gaussian basis set \citep{porezag_optimization_1999}, numerically precise
variational integration and an analytic solution of Poisson’s equation to accurately determine the self-consistent potentials, secular matrix, total energies and Hellmann – Feynman – Pulay forces. The
exchange correlation is modeled by GGA \citep{ perdew_accurate_1992,perdew_generalized_1996-1} in the form of Perdew-Burke-Enzerhof (PBE)
\citep{perdew_generalized_1996}. The relaxation was terminated once forces below 0.05 eV/$\mathring{\text{A}}$ per atom have been reached. Energetically favored structures are now subject to a stability
analysis. First of all the the binding energy per atom E$_b$ of the cluster (Co$_{\text{n}}$Mo$_{\text{m}}$) is considered which is defined as: 
 \begin{equation}
   \text{E}_b = \frac{  \text{E} (\text{Co}_{\text{n}}\text{Mo}_\text{m})  - \text{n} \cdot  \text{E}(\text{Co})  - \text{m} \cdot  \text{E}(\text{Mo}) }{\text{m+n}}
 \end{equation}
Here E$_b < 0$ refers to a situation where the total energy of the given cluster is smaller than the sum of its parts, hence the system can save energy by clustering up. For E$_b > 0$ on the other
hand one would expect a separation of the cluster into smaller components. Note that it is not always comprehensible from the literature cited in this paper if a binding energy is
given per atom or not. Once a cluster proved energetically stable we computed the vibrational spectra to check further for dynamical stability. Unstable clusters will show imaginary frequencies. \\
As some of the stable clusters also showed a remaining magnetic moment S we computed the magnetic anisotropy D. The magnetic anisotropy is mainly due to spin-orbit coupling \citep{vleck_1937} and can be obtained within DFT via second order perturbation theory \citep{pederson_1999,perderson_1999a}:
\begin{equation} \label{second_order}
 \Delta_2\,=\,\sum_{\sigma\sigma'}\sum_{ij}M_{ij}^{\sigma\sigma'}S_{i}^{\sigma\sigma'}S_{j}^{\sigma'\sigma},
\end{equation}
where $\Delta_2$ is the second order perturbation energy, $\sigma$ denotes different spin degrees of freedom and i,j are coordinate labels $x,y,z$. Within this framework
$S_{i}^{\sigma\sigma'}$ is defined as 
\begin{equation}
 S_{i}^{\sigma\sigma'}\,=\,\langle \chi^{\sigma}\mid S_i\mid\chi^{\sigma'}\rangle,
\end{equation}
where $\chi^{\sigma}$ and $\chi^{\sigma'}$ are a set of spinors. These spinors are constructed from a unitary transformation of the $S_z$ eigenstates. The matrix element $M_{ij}^{\sigma\sigma'}$ is
given by:
\begin{equation} \label{level_dependence}
 M_{ij}^{\sigma\sigma'}\,=\,-\sum_{kl} \frac{\langle\varphi_{l\sigma} \mid \hat{V}_i \mid \varphi_{k\sigma'}\rangle\langle\varphi_{k\sigma'} \mid \hat{V}_j \mid
\varphi_{l\sigma}\rangle}{\epsilon_{l\sigma}-\epsilon_{k\sigma'}}
\end{equation}
with the occupied and unoccupied states $\varphi_{l\sigma}$ and $\varphi_{k\sigma'}$ and the respective energies $\epsilon_{l\sigma}$ and $\epsilon_{k\sigma'}$.  This method can be applied to
molecules of arbitrary symmetry and has been used successfully for the prediction of the magnetic anisotropy of various single molecule magnets \citep{kort2003,post2006}. In the absence of a magnetic
field the second order perturbation energy can be rewritten in terms of the anisotropy tensor $D_{ij}$: 
\begin{equation}
 \Delta_2\,=\,\sum_{ij}D_{ij} \langle S_i \rangle \langle S_j \rangle.
\end{equation}
For a diagonal form of the $D$ tensor the following expression is obtained: 
\begin{equation}
D\,=\,D_{zz}-\frac{1}{2}(D_{xx}+D_{yy}) 
\end{equation}
Within this framework D$<0$ refers to an easy axis behavior and $D>0$ indicates an easy plane system. For an easy axis system the anisotropy barrier U is given by U\,=\,S$^2|$D$|$. Additionally we
apply the correction proposed by Van W\"ullen \citep{wuel2009}. \\
For comparison we have also computed the ionization potential IP\,=\,E(N-1) - E(N) as well as the electron affinity EA\,=\,E(N) - E(N+1) where N is the total number of electrons in the system.
These two quantities are also closely related to the chemical reactivity which can be described in terms of the chemical potential $\mu$ and the molecular hardness $\eta$ \citep{parr1983,parr1984}.
The chemical potential is a measure of the escaping tendency of electrons from a cluster and is defined as:
\begin{equation}
 \mu = -\frac{1}{2} (\text{IP+EA}).\label{potential}
\end{equation}
The molecular hardness on the other side is given by: 
\begin{equation}
 \eta = \frac{1}{2} (\text{IP-EA}) \label{hardness}
\end{equation}
and accounts for the resistance of the chemical potential to a change in the number of electrons, i.e. it is related to the reactivity of the cluster. Note that the hardness is also related to the
HOMO-LUMO gap. A small hardness indicates a small gap and therefore we expect an increase in the reactivity. An other interesting property related to quantum mixing is the optical polarizability. Here
a small gap/molecular hardness leads to a stronger mixing and hence a larger polarizability. A further important quantity is the absolute electronegativity defined as $\chi = -\mu$, where large
$\chi$ values characterize acids and small $\chi$ values characterize bases. 
\\
Furthermore it has been shown \citep{naele1963} that the bond dissociation energy D$_0$ is related to the electron affinity and the ionization potential via: 
\begin{equation}
 D_0 = (\text{IP-EA})
\end{equation}
Although the bond dissociation energy and the binding energy are different quantities they are sometimes used misleadingly as equivalent in literature. 
Note that up to m+n\,=\,3 there is always only one conformation for each stoichiometric composition. For larger clusters the number of different variations grows very fast, hence only the best will be
presented in the present paper. All images of clusters are created using Jmol \citep{jmol}.

\section{Results}

\renewcommand{\arraystretch}{1.1}

\begin{table*} [t]
\setlength{\tabcolsep}{8pt}
 \centering
 \caption{Binding energies E$_b$ in eV, magnetic ground state S, magnetic anisotropy D, electron affinity (EA) in eV, ionization potential (IP) in eV, the chemical potential $\mu$ in eV, the
molecular hardness $\eta$ in eV, the gap between the highest occupied molecular orbital (HOMO) and the lowest unoccupied molecular orbital (LUMO) in eV as well as the average bond
distance for the possible Co$_n$Mo$_m$ clusters}
\label{overview}
\begin{tabular}{c|cc|cccccc|ccc}
%\begin{tabular}{c|cc:cccccc:ccc}
Cluster  	& S 		& D 	& E$_b$		&EA	& IP	& $\mu$	& $\eta$& Gap	& d$_{Co-Co} $ 		& d$_{Co-Mo} $ 		& d$_{Mo-Mo}$ \\ 	
		&  		& [K]  	&  [eV] 	&[eV] 	&[eV] 	&[eV]	&[eV]	& [eV]	&[$\mathring{\text{A}}$]& [$\mathring{\text{A}}$]&[$\mathring{\text{A}}$]	 \\  
\hline\noalign{\smallskip}
Co$_2$ 	 	& 2		& -5.6	& 1.29 		& 0.8	& 7.3	& -4.02	& 3.22	&	& 1.99 			&- 			& - \\                    
CoMo 	 	& $\frac{3}{2}$	& 13.4	& 0.83 		& 0.6	& 6.7	& -3.68	& 3.05	& 0.58	& - 			&2.48 			& - \\            	
Mo$_2$ 	 	& 0	 	& -	& 1.51 		& 0.5	& 6.7	& -3.63	& 3.07	& 1.17	&- 			&- 			&1.98 \\                 	
 \hline     \noalign{\smallskip}                                                                                                                                                                     
Co$_3$ 	 	& $\frac{5}{2}$	& -6.5	& 1.65 		& 1.4	& 6.0	& -3.71 & 2.32	& 0.07	& 2.21			&- 			& - \\                			
Co$_2$Mo 	& 2		& 12.5	& 1.31 		& 1.3	& 6.2	& -3.80 & 2.36	& 0.28	& 2.10			& 2.48 			& - \\               			
CoMo$_2$ 	& $\frac{1}{2}$ & -	& 1.81		& 0.8	& 5.9	& -3.36 & 2.58	& 0.63	& - 			& 2.47			& 2.07 \\            			
Mo$_3$ 	 	& 0		& -	& 1.72 		& 0.7	& 5.7	& -3.26 & 2.50	& 0.61	&- 			&- 			& 2.31	 \\      		
 \hline    \noalign{\smallskip}                                                                                                                                                                         
Co$_4$ 	 	& 5		& 1.0	& 2.09		& 1.3	& 6.1	& -3.72 & 2.25	& 0.27	& 	2.31		&	-		&-	\\           	
Co$_3$Mo 	& $\frac{5}{2}$	& 10.7	& 1.74		& 1.3	& 6.1	& -3.70 & 2.39	& 0.23	&	2.29		&	2.39		&-	\\        		
Co$_2$Mo$_2$ 	& 2		& -28.8	& 1.98		& 1.3	& 5.0	& -3.21 & 1.85	& 0.28	&	2.58		&	2.40		& 2.12	\\              			
CoMo$_3$ 	& $\frac{3}{2}$	& 8.5	& 2.08		& 1.0	& 5.6	& -3.29 & 2.32	& 0.49	&	-		&	2.43		& 2.42	\\       	
Mo$_4$ 		& 0		& -	& 2.25		& 0.7	& 6.7	& -3.74 & 2.98	& 0.75	&	 -		& 	-		& 2.52	\\		
 \hline     \noalign{\smallskip}                                                                                                                                                                      
Co$_5$ 		& $\frac{11}{2}$& -1.2	& 2.36		& 1.6	& 6.1	& -3.90 & 2.26	& 0.22	& 2.34			&	-		& -	\\	
Co$_4$Mo 	& 4		& -1.5	& 2.27		& 1.1	& 5.9	& -3.48 & 2.39	& 0.27	& 2.29			&	2.32		& -	\\	
Co$_3$Mo$_2$ 	& $\frac{5}{2}$	& -1.6	& 2.36		& 1.4	& 6.2	& -3.80 & 2.39	& 0.23	& 2.32			&	2.38		& 2.17	\\	
Co$_2$Mo$_3$ 	& 2		& 6.1	& 2.35		& 1.2	& 5.8	& -3.52 & 2.28	& 0.48	& 2.21			&	2.47		& 2.35	\\	
Co$_1$Mo$_4$ 	& $\frac{1}{2}$	& -	& 2.41		& 0.9	& 5.2	& -3.09 & 2.11	& 0.28	& - 			&	2.46		& 2.55	\\	
Co$_5$ 		& 0		& -	& 2.47		& 0.7	& 5.2	& -2.98 & 2.23	& 0.58	& -			&	-		& 2.53	\\	
 \hline    \noalign{\smallskip}                                                                                                                                                                      
Co$_6$ 		& 7		& -0.02	& 2.73		& 1.5	& 6.6	& -4.02	& 2.54	& 0.45	& 2.30			&	-		& -	\\		
Co$_5$Mo 	& $\frac{9}{2}$	& -0.6	& 2.57		& 1.4	& 5.9	& -3.63	& 2.23	& 0.22	& 2.32			&	2.37		& -	\\		
Co$_4$Mo$_2$	& 3		& -2.7	& 2.66		& 1.5	& 5.7	& -3.56	& 2.09	& 0.27	& 2.33			&	2.41		& 2.25	\\		
Co$_3$Mo$_3$	& $\frac{5}{2}$	& 3.6	& 2.72		& 1.6	& 5.7	& -3.61	& 2.06	& 0.37	& 2.33			&	2.42		& 2.40	\\		
Co$_2$Mo$_4$	&	2	& -10.8	& 2.72		& 0.5	& 5.6	& -3.05	& 2.56	& 0.46	& 2.26			&	2.49		& 2.45	\\		
CoMo$_5$ 	& $\frac{1}{2}$	& -	& 2.74		& 1.1	& 5.7	& -3.41	& 2.33	& 0.53	& -			&	 2.59		& 2.50	\\		
Mo$_6$	 	& 0		& -	& 2.84		& 0.1	& 5.2	& -2.67	& 2.56	& 0.73	& - 			& -			& 2.55	\\		
\end{tabular}
 \end{table*}

\subsection{Dimers}

The smallest possible clusters are dimers and within the dimers only three different compositions for nano\-clusters are possible as shown in table \ref{overview}. The pure Co dimer, the pure Mo dimer
and a mixed CoMo cluster. Within the present study the Mo dimer exhibits a non magnetic ground state (S\,=\,0) and a binding energy of E$_b$\,=\,1.5\,eV.  There has been extensive experimental as well as
theoretical research on the Mo dimer. A very nice overview is given by Diez \citep{diez2000}, Zhang et al \citep{zhang2004} or more recently by Aguilera-Granja et al \citep{gran2008}. There are
reports
of calculated binding energies ranging from 1.36 - 2.67\,eV (see \citep{zhang2004} and references therein) depending on the level of approximation used. Older works report even lower binding energies
well below 1 eV \citep{good1982, atha1982}. The experimental value for the binding energies $ \sim $ 2.2\,eV \citep{efremov1978,morse1986,simard1998}. Within the present work the equilibrium distance of
the Mo atoms is 1.98\,$\mathring{\text{A}}$, which is in good agreement with various theoretical (d$_{Mo-Mo} \sim $ 1.8-2.1\,$\mathring{\text{A}}$ \citep{zhang2004,diez2000} and references therein) and
experimental (d$_{Mo-Mo} \sim $ 1.94\,$\mathring{\text{A}}$ \citep{efremov1978,hopkins1983,morse1986,simard1998}) results. We observe dissociation energy D$_0$ of 6.2\,eV, which is slightly larger than
those previously reported (D$_0$\,=\,4.5\,eV \citep{simard1998}- 5\,eV\citep{andzelm1985}). Furthermore we computed a vibrational frequency $\omega$\,=\,533 cm$^{-1}$, which is well in line with other
theoretical works ($\omega$\,=\,360-552 cm$^{-1}$, see \citep{diez2000} and references therein) and reasonably close to the experimental value of $\omega$\,=\,477cm$^{-1}$, see\citep{diLella1982,morse1986}.
The ionization potential of 6.7\,eV is in good agreement with experimental studies (6.4\,eV $\le$ IP $\le$ 8.0; see \citep{simard1998} and references therein).\\
For the Co$_2$ dimer we calculate a ferromagnetic ground state S\,=\,2 at an equilibrium distance d$_{Co-Co}$\,=\,1.98\,$\mathring{\text{A}}$ which is consistent with previously reported theoretical
\citep{castro_structure_1997,fan_geometry_1997,jamorski_structure_1997,pereiro2001,wang_low-lying_2005,strandberg_transition-metal_2007,datta2007,sebetci_cobalt_2008,fritsch_transition_2008,
strandberg_calculation_2008} and experimental\citep{gambardella_giant_2003} values. The S\,=\,2 ground state indicates a s$^1$d$^8$ electronic configuration at each Co atom which is indeed energetically
favorable compared to the s$^2$d$^7$ electronic configuration \citep{jamorski_structure_1997,fan_geometry_1997}. Additionally we report an easy axis magnetic anisotropy. We note that the strength of
the anisotropy depends crucially on the bonding distance and therefore also on the energetic ground state. This is in line with the results of Wang et al. \citep{wang_low-lying_2005}, who reported low
lying quintet states for the Co dimer. This makes it far from trivial to get the energetic ground state of Co$_2$ correctly. The predicted ground state and hence electronic structure therefore depends
crucially on the exact details (level of approximation for the exchange-correlation functional, convergence criteria, etc.) of the calculation \citep{wang_low-lying_2005,ma2006,yanagisawa2000,fritsch_transition_2008}. As shown in equation (\ref{level_dependence}) the computation of D depends crucially on the electronic structure, hence it is
not surprising that we find strong changes in D depending on the details of the calculation. Furthermore we report a binding energy of E$_b$\,=\,1.3\,eV which is within the range of previously reported
theoretical values (0.87 - 5.4\,eV (see \citep{sebetci_cobalt_2008} and references therein), and very close to the experimental value of E$_b \le 1.4$\,eV \citep{russon1994, hales1994}. The ionization
potential of 7.29\,eV is also in good agreement with other experimental (IP $\sim$ 6.3\,eV) \citep{morse1986,parks1990,hales1994} and theoretical (IP\,=\,6.8 - 7.5\,eV)\citep{jamorski_structure_1997,pereiro2001,sebetci_cobalt_2008} reports. The electron affinity on the other hand is with only 0.8\,eV very small. Accordingly we get a dissociation energy of 5.5\,eV,
which is larger compared to previously reported values (3.1\,$\le$ D$_0$ $\le$\,4.8) \citep{calaminici_density_2007}. The computed vibrational frequency of $\omega$\,=\,380\,cm$^{-1}$ is comparable to
other
GGA calculations ($\omega$\,=\,340-420 cm$^{-1}$) \citep{pereiro2001,castro_structure_1997,jamorski_structure_1997,calaminici_density_2007}, whereas the experimental frequency is given by 290\,cm$^{-1}$, see\citep{diLella1982}.\\
The CoMo dimer is stable with respect to the vibrational spectra as well as binding energy (E$_b$\,=\,0.8\,eV) and shows a $S\,=\,\frac{3}{2}$ magnetic ground state with an easy plane magnetic anisotropy. However the binding energy is a factor of two smaller compared to the pure Mo$_2$ and Co$_2$ resulting in a much larger bonding distance of d$_{Co-Mo} $\,=\,2.48\,$\mathring{\text{A}}$.
Furthermore we report an easy axis anisotropy of 13.4\,K. Additionally we predict a pronounced Raman feature at 202 cm$^{-1}$. \\
In general we see an decrease in the EA as well as in the IP with increasing Mo concentration. This leads to an increase in the chemical potential $\mu$ as well as an increase in the molecular
hardness $\eta$ with increasing Co content with Co$_2$ showing the largest $\mu$\,=\,-4.0\,eV and $\eta$\,=\,3.2\,eV. Accordingly Co$_2$ is the most stable and least polarizable candidate out of these three dimers.

\subsection{Trimers}

\begin{figure}[h]
  \centering
 \subfloat [] {
 \includegraphics[scale=0.2]{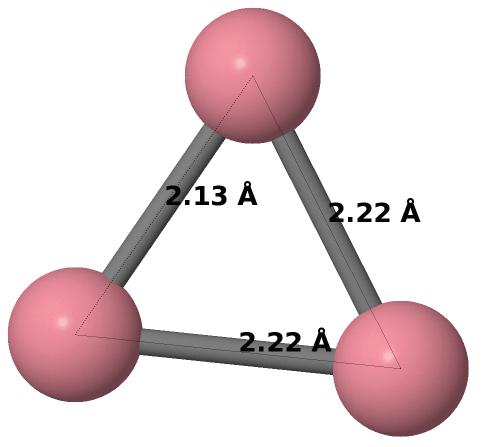}
 \label{co3}
}
 \subfloat [] {
 \includegraphics[scale=0.2]{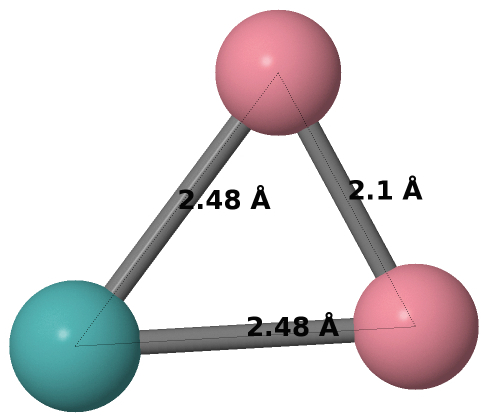}
 \label{co2mo}
}
\\
 \subfloat [] {
 \includegraphics[scale=0.2]{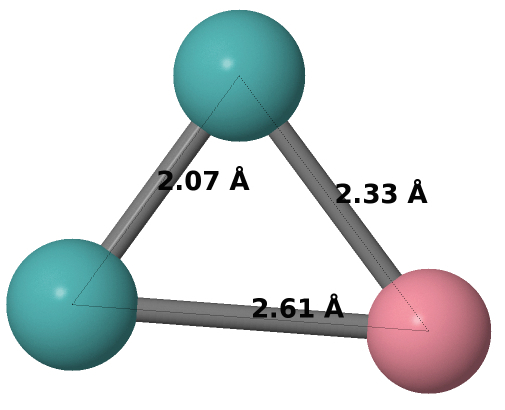}
 \label{como2}
}
 \subfloat [] {
 \includegraphics[scale=0.2]{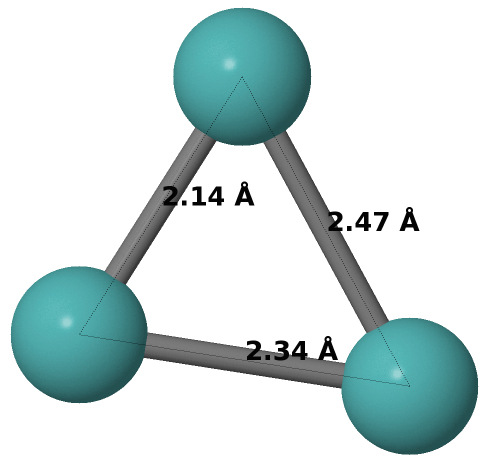}
 \label{mo3}
}
 \caption[]{a): Co$_3$, b): Co$_2$Mo, c): CoMo$_2$, d): Mo$_3$; blue: Mo, red: Co}
 \label{trimer}
 \end{figure}

There are four different compositions for the trimers all of which are found to be stable as shown in table \ref{overview} and depicted in figure \ref{trimer}. As we start from the finite sphere packing all of
them form a triangular structure as shown in figure \ref{trimer}.

 The Mo$_3$ cluster shows again a non magnetic ground state with an increased binding energy E$_b$\,=\,1.7\,eV and an increased average inter-molecular distance d\,=\,2.31\,$\mathring{\text{A}}$ compared
to the respective dimer. This includes one short bond of 2.14\,$\mathring{\text{A}}$ and two longer bonds (2.34\,$\mathring{\text{A}}$ and 2.47\,$\mathring{\text{A}}$), which is in line with previously
reported work
\citep{berces1998,diez2000,zhang2004,Koteski200579} and shows the tendency of Mo cluster for dimerization \citep{zhang2004,gran2008} due to the half-filled shell of the Mo atom. Furthermore we observe
an increase in the electron affinity by 0.2\,eV compared to the dimer as well as an decrease in the ionization potential by 1\,eV. This leads consequently to a lower chemical potential ($\mu$\,=\,-3.2\,eV)
and molecular hardness ($\eta$\,=\,2.5\,eV). Note that the magnetization, binding energy and the inter-molecular distance depends crucially on the symmetry of the Mo triangle. While some studies report a
magnetic moment of 0.67\,$\mu_B$/atom others predict a magnetic moment of 0\,$\mu_B$/atom \citep{berces1998,diez2000,zhang2004}. For the structure found in our calculations we predict pronounced Raman frequencies at
66, 220 and 406\,cm$^{-1}$. \\
For the Co$_3$ cluster we found a S\,=\,$\frac{5}{2}$ ground state with an average distance of d$_{Co-Co} $\,=\,2.21\,$\mathring{\text{A}}$ and a binding energy of 1.6\,eV/atom, which is in good agreement with experimental data (E$_b \ge$ 1.5\,eV)\citep{hales1994}. There is an ongoing discussion in literature depending the energetically favored alignment of three Co atoms. Sebetci \citep{sebetci_cobalt_2008}
and Ma et al \citep{ma2006} predict a linear structure to be the most stable one, whereas several other authors \citep{jamorski_structure_1997,fan_geometry_1997,
castro_structure_1997,pereiro2001,datta2007,ganguly2008} find a triangular structure to be the ground state. The triangular structure is also observed experimentally \citep{VanZee1992}. The bond
length
for the triangular structure varies from 2.04 - 2.24\,$\mathring{\text{A}}$ (see \citep{sebetci_cobalt_2008} and references therein) and agrees well with the bond length computed in the present paper.
The binding energy on the other hand ranges from 1.70\,eV to 5.34\,eV (see \citep{sebetci_cobalt_2008} and references therein) where the E$_b$ computed in this paper presents an lower bound. The reported
ground states for the Co trimer are S\,=\,$\frac{7}{2}$ (see Ref.~[\onlinecite{pereiro2001,fan_geometry_1997,ma2006,sebetci_cobalt_2008}]) and S\,=\,$\frac{5}{2}$ (see Ref.~[\onlinecite{jamorski_structure_1997,castro_structure_1997,datta2007,ganguly2008}]), which is consistent with the experimental report of Zee et al. \citep{VanZee1992} who could not clearly distinguish between
S\,=\,$\frac{7}{2}$ and  S\,=\,$\frac{5}{2}$. The same behavior is observed by Ganguly et al \citep{ganguly2008}, who found two degenerated ground states of S\,=\,$\frac{5}{2}$ and S\,=\,$\frac{7}{2}$. In the present study we also observe a structure with S\,=\,$\frac{7}{2}$ which is only 4.5 meV higher in energy than the actual ground state of S\,=\,$\frac{5}{2}$. The magnetic ground state depends therefore crucially on the actual geometry (i.e. bond lenght, angles, etc.) of the structure. Furthermore we report an easy axis magnetic anisotropy for the Co trimer resulting in a barrier of 41\,K. The
ionization potential of 6.0\,eV is in good agreement with experimental (IP\,=\,5.97\,eV) \citep{yang1990} and theoretical (IP\,=\,6.6\,eV) \citep{jamorski_structure_1997} results. The dissociation energy of 1.55\,eV/atom is also in good agreement with literature (D$_0$\,=\,1.45\,eV) \citep{perez_physical_2012}, although they predict a much smaller IP\,=\,4.7\,eV than the one observed here and in experiment.
Consequently a larger electron affinity is computed by Perez et al \citep{perez_physical_2012} (EA\,=\,2.6\,eV) compared to the one observed here (EA=1.4\,eV). As already observed for the Mo$_3$ there is
again an increase in the electron affinity (+0.6\,eV) compared to the dimer as well as an decrease in the ionization potential (-1.3\,eV). This leads consequently to a lower chemical potential ($\mu$\,=\,-3.7\,eV) and molecular hardness ($\eta$\,=\,2.3\,eV). Therefore we expect the Co$_3$ cluster to be more polarizable and less acidic compared to Co$_2$. \\
For the Co$_2$Mo trimer we obtain an decreased Co-Co distance of 2.1\,$\mathring{\text{A}}$ compared to the Co-Co distance in the Co$_3$ trimer. The Co-Mo distance (2.48\,$\mathring{\text{A}}$) on the
other hand remains constant compared to the respective distance in the CoMo dimer (2.48\,$\mathring{\text{A}}$). The Co$_2$Mo trimer exhibits a magnetic ground state of S\,=\,2 with an easy plane
anisotropy of 12.5\,K.  The EA (\,=\,1.3\,eV) and IP (\,=\, 6.2\,eV) are closely related to those of Co$_3$ which leads to comparable chemical reactivity in terms of $\mu$ and $\eta$. Furthermore we predict
pronounced Raman frequencies at 147, 186 and 337\,cm$^{-1}$.\\
The CoMo$_2$ trimer on the other hand resembles more the Mo trimer with respect to the binding energy, inter-atomic distances and EA as well as IP (see table \ref{overview}) which is consistent with
the amount of Mo within the trimer. Again we observe one short Mo-Mo distance of 2.07 $\mathring{\text{A}}$ which indicates a dimerization of Mo. We report an S\,=\,$\frac{1}{2}$ ground state and expect
pronounced Raman frequencies at 104, 236 and 388\,cm$^{-1}$. \\

\subsection{Tetramers}
\begin{figure}[h]
  \centering
 \subfloat [] {
 \includegraphics[scale=0.2]{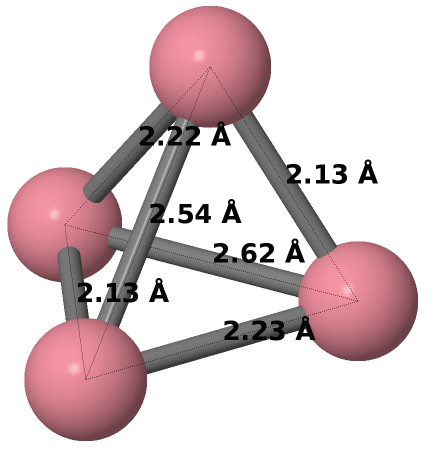}
\label{co4}
}
 \subfloat [] {
 \includegraphics[scale=0.2]{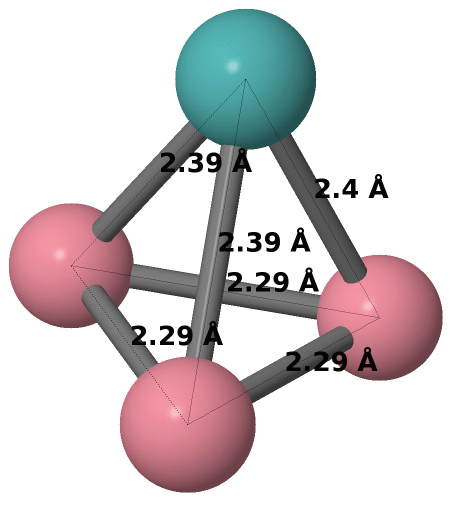}
 \label{co3mo}
}
\\
 \subfloat [] {
 \includegraphics[scale=0.2]{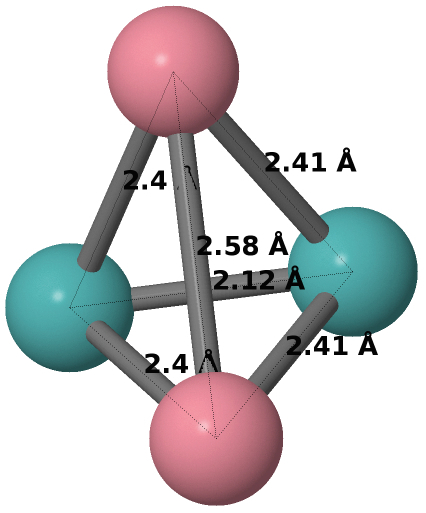}
 \label{co2mo2}
}
 \subfloat [] {
 \includegraphics[scale=0.2]{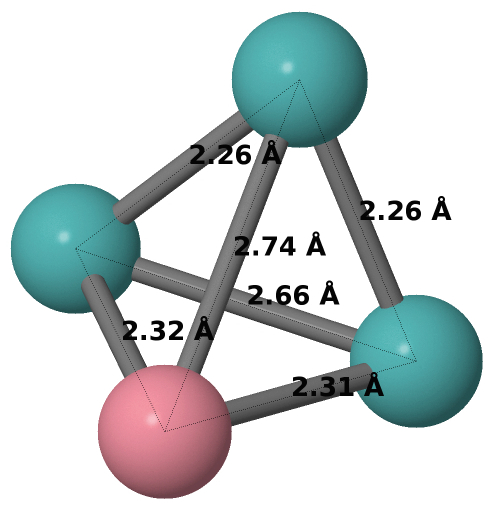}
 \label{como3}
}
\\
 \subfloat [] {
 \includegraphics[scale=0.2]{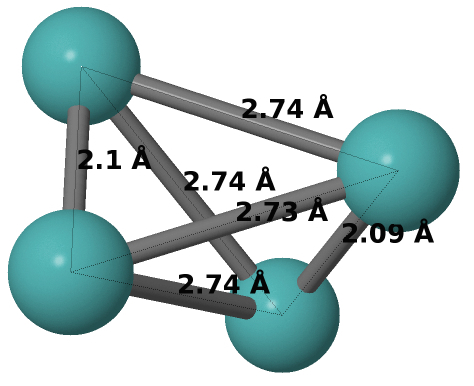}
 \label{mo4}
}
 \caption[]{a): Co$_4$, b): Co$_3$Mo, c): Co$_2$Mo$_2$, d): CoMo$_3$, e): Mo$_4$; blue: Mo, red: Co}
 \label{tetramer}
 \end{figure}

The calculated ground states for the different tetramers are summarized in table \ref{overview} and depicted in figure \ref{tetramer}. As we start from the finite sphere packing only tetrahedral and
no square planar structures are considered as initial geometry for the optimization. However the optimization is unconstrained so it is possible for the structures to relax to a square planar
structure.

For the Mo$_4$ cluster we observe a non magnetic, flattened tetrahedral ground state with a binding energy of 2.25\,eV/atom and two different distances of 2.73\,$\mathring{\text{A}}$ and 2.10\,$\mathring{\text{A}}$. There are two short distances with is in line with the already mentioned dimerization tendency of Mo clusters \citep{zhang2004,gran2008}. The ionization potential and the
electron affinity are nearly identical with those of the dimer. The structure is also stable with respect to Raman frequencies where we predict pronounced peaks at 145, 163, 195, 435 and 444\,cm$^{-1}$.  Few studies on the Mo tetramer are available in literature. Min et al \citep{min2001} report slightly flattened tetrahedron as the ground state configuration with a binding energy of 3
eV/atom and four equivalent bonds of 2.31\,$\mathring{\text{A}}$ and two elongated bonds of 2.63\,$\mathring{\text{A}}$ and 2.75\,$\mathring{\text{A}}$ length. Diez \citep{diez2000} predict a ground
state
structure with a binding energy of 2.59\,eV/atom and bond lengths of 3.00\,$\mathring{\text{A}}$ and 2.23\,$\mathring{\text{A}}$ as well as a magnetic ground state of S\,=\,0. Energetically close to that structure they observed a second non magnetic structure with a binding energy of 2.57\,eV/atom and bond lengths of 2.62\,$\mathring{\text{A}}$ and 2.12\,$\mathring{\text{A}}$. They predict vibrational frequencies of 147, 195, 213, 437 and 459\,cm$^{-1}$, which is in very good agreement to the frequencies given in this work. Additionally they observed a variety of structures close to the ground
state with differing magnetic ground states. Zhang et al. \citep{zhang2004} and Aguilera-Granja et al \citep{gran2008} on the other hand predict a rhombic ground state structure with an S\,=\,2
\citep{gran2008} or S\,=\,0 \citep{zhang2004} magnetic ground state. However both studies find a isomer close in energy that is nearly identical to the structure reported here and by others
\citep{diez2000,min2001}. \\
The Co$_4$ cluster on the other hand exhibits a binding energy of 2.1\,eV/atom with an equilibrium distance of 2.31\,$\mathring{\text{A}}$. This includes four short bonds of length 2.17\,$\mathring{\text{A}}$ and two long bonds length 2.77\,$\mathring{\text{A}}$ forming again a flattened tetrahedron, which is in line with previous studies
\citep{li_electronic-structure_1993,castro_structure_1997,andriotis1998,jamorski_structure_1997,pereiro2001_a,lopez2003,datta2007,ganguly2008,perez_physical_2012}. Other studies
\citep{fan_geometry_1997,ma2006,sebetci_cobalt_2008} predict a slightly out-of plane rhombus to be the ground state. We get a total magnetic moment of S\,=\,5 in agreement with previous studies
\citep{jamorski_structure_1997,castro_structure_1997,fan_geometry_1997,guevara1997,andriotis1998,ganguly2008,sebetci_cobalt_2008,perez_physical_2012}. There are also reports of S\,=\,8 (see Ref.~[\onlinecite{ma2006}]) and S\,=\,9 (see Ref.~[\onlinecite{li_electronic-structure_1993}]). Those studies found the S\,=\,5 ground state to be slightly larger in energy compared to the respective magnetic ground states reported there. We also found
an
easy plane magnetic anisotropy of 1.0\,K. Furthermore we report an ionization potential of 6.1\,eV, which is in very good agreement with experimental data (IP $\sim$ 6.2\,eV)
\citep{yang1990,parks1990,hales1994} and reasonably close to other theoretical works (IP\,=\,5.5-5.7\,eV) \citep{guevara1997, perez_physical_2012}. We also found pronounced Raman features at
69, 145, 211, 286 and 359\,cm$^{-1}$, which is in agreement with previous theoretical work \citep{andriotis1998,castro_structure_1997,jamorski_structure_1997,sebetci_cobalt_2008}.\\
For the mixed cluster we observe an increase in the total magnetic moment as well as a steady decrease in the binding energy with increasing Co content. For CoMo$_3$ we found a magnetic ground
state of S\,=\,$\frac{3}{2}$ with an easy plane anisotropy of 10.7\,K, a binding energy of 2.0\,eV, a slightly larger electron affinity (0.9\,eV) and a much smaller ionization potential (5.6\,eV) compared to
the Mo tetramer. Furthermore we report pronounced Raman features at 73, 118, 160, 182, 276 and 363\,cm$^{-1}$. For Co$_2$Mo$_2$ we predict a S\,=\,2 magnetic ground state with a very large easy axis
anisotropy of -28.8\,K and a binding energy of 1.9\,eV. The electron affinity (1.3\,eV) goes again up compared to the Mo tetramer and is approximately equal to the electron affinity of a pure Co
tetramer. The ionization potential (5\,eV) on the other hand is is extremely low in comparison to the other tetramers. This results in a very low hardness ($\eta$\,=\,1.8\,eV) which makes this tetramer
easily polarizable. For Co$_3$Mo we found an even higher magnetic ground state of S\,=\,$\frac{5}{2}$, an easy plane anisotropy of 8.5\,K and a binding energy of 1.7\,eV. The electron affinity (1.3\,eV) and
the ionization potential (6.1\,eV) is very close to the respective values computed for Co$_4$. Respectively we expect a chemical behavior ($\mu, \eta$) nearly identical to the Co tetramer. Furthermore
we report pronounced Raman features at 121, 194, 225 and 342\,cm$^{-1}$. \\

\subsection{Pentamers}
\begin{figure}[h]
  \centering
 \subfloat [] {
 \includegraphics[scale=0.2]{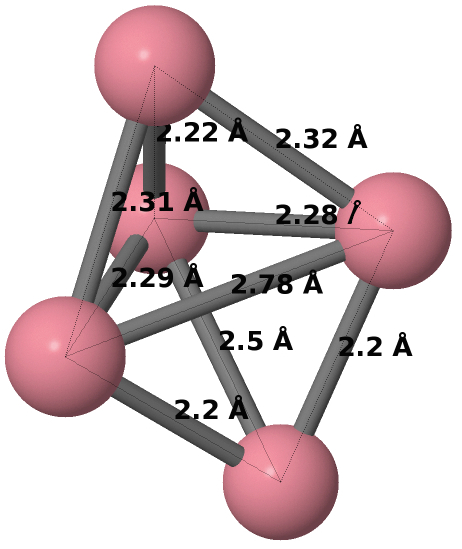}
\label{co5}
}
 \subfloat [] {
 \includegraphics[scale=0.2]{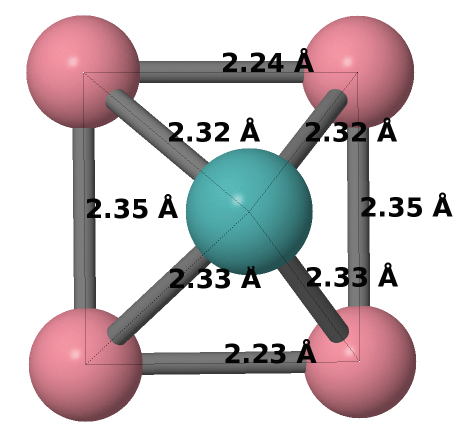}
 \label{co4mo}
}
\\
 \subfloat [] {
 \includegraphics[scale=0.2]{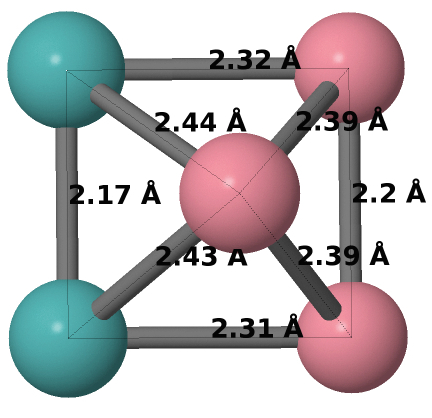}
 \label{co3mo2}
}
 \subfloat [] {
 \includegraphics[scale=0.2]{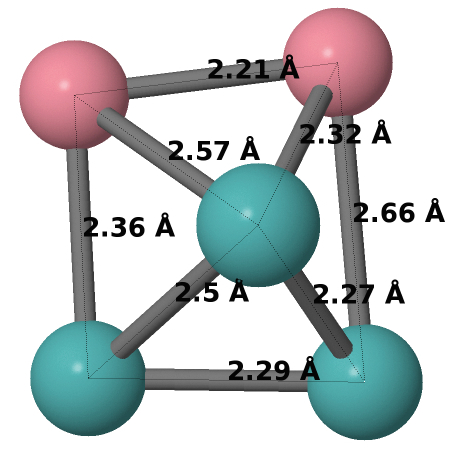}
 \label{co2mo3}
}
\\
 \subfloat [] {
 \includegraphics[scale=0.2]{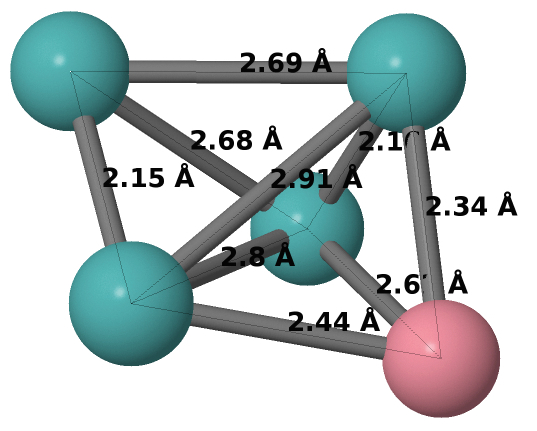}
 \label{como4}
}
 \subfloat [] {
 \includegraphics[scale=0.2]{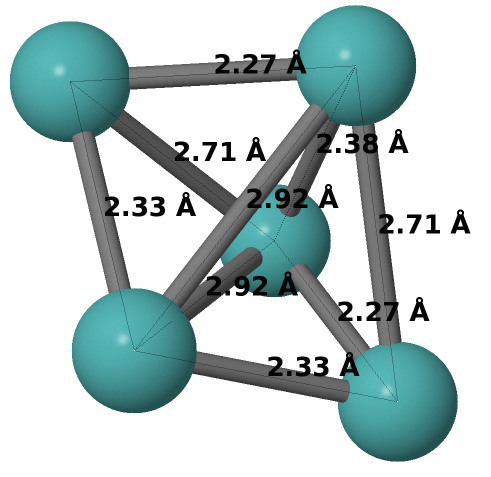}
 \label{mo5}
}
 \caption[]{a): Co$_5$, b): Co$_4$Mo, c): Co$_3$Mo$_2$, d): Co$_2$Mo$_3$, e): CoMo$_4$, f): Mo$_5$; blue: Mo, red: Co}
 \label{pentamer}
 \end{figure}

For the pentamer structure there are two different possible structures - square pyramidal (7 possible bonds) and trigonal bypiramidal (9 possible bonds) - and six stoichiometric
compositions. The respective ground state structure and selected properties of those structures are summarized in table \ref{overview}, whereas the ground state geometry can be found in figure
\ref{pentamer}. 

 For the Mo$_5$ cluster we observe a distorted trigonal bipyramidal ground state structure (which could also be viewed as a distorted square pyramid) with no magnetic moment and a binding energy of 2.5\,eV. The average bonding distance is 2.53\,$\mathring{\text{A}}$, with no exceptional short Mo-Mo distances. Hence the tendency to form dimers seems to be weakened in the pure Mo pentamer which was also
observed by others \citep{zhang2004}. There is no change in the electron affinity compared to smaller Mo clusters whereas the ionization potential (5.2\,eV) is considerably lowered. Consequently we
observe a reduced chemical potential $\mu$ and a reduced hardness $\eta$ compared to smaller clusters. Furthermore we predict pronounced Raman features at 138, 154, 169, 207, 229, 290 and 375\,cm$^{-1}$. Koteski et al \citep{Koteski200579} reported a similar ground state structure. Zhang et al. \citep{zhang2004} report trigonal bipyramidal ground state with S\,=\,0 with six short distances of 2.26\,$\mathring{\text{A}}$ and three long distances of length 2.77\,$\mathring{\text{A}}$. Very close in energy (E$_b$\,=\,+0.02\,eV/atom compared to the trigonal bipyramidal structure) they found a pyramid with a rectangular base with bond length of 2*1.94\,$\mathring{\text{A}}$, 2*2.94\,$\mathring{\text{A}}$ and 4*2.48\,$\mathring{\text{A}}$. This structure is also non magnetic. Nearly identical results are reported by Min et al \citep{min2001}. Aguilera-Granja et al \citep{gran2008} on the other hand predict a 2D fan-like ground state with an S\,=\,4 magnetic ground state. This structure was also proposed by Zhang et al. \citep{zhang2004} however they reported a binding energy well below the one of the ground state. \\
For Co$_5$ we observe the same ground state structure as already mentioned for Mo$_5$ a distorted trigonal bipyramid or a distorted square pyramid. The structure has a net
magnetic moment of S\,=\,$\frac{11}{2}$ with an easy axis anisotropy leading to a barrier of 36\,K and a binding energy of 2.36\,eV/atom. The average Co-Co distance 2.34\,$\mathring{\text{A}}$ with bonds
ranging from 2.2\,$\mathring{\text{A}}$ - 2.78\,$\mathring{\text{A}}$ as depicted in figure \ref{co5}. A very similar structure with S\,=\,$\frac{13}{2}$ was found only 0.08 eV above the ground state
structure. Furthermore we expect pronounced Raman features at 92, 116, 133, 182, 204, 249, 300 and 334\,cm$^{-1}$. There is no consistent picture in the literature concerning the ground state geometry of Co$_5$. Sebetci et al \citep{sebetci_cobalt_2008} report a fan-like 2d structure with S\,=\,$\frac{11}{2}$ to be the ground state with the energetically close states of a trigonal bipyramidal
(E$_b$\,=\,+0.03\,eV/atom, S\,=\,$\frac{13}{2}$) and square pyramidal (E$_b$\,=\,+0.04\,eV/atom, S\,=\,$\frac{11}{2}$) structure. Other studies report a square pyramidal \citep{fan_geometry_1997,guevara1997,andriotis1998} or trigonal bipyramidal \citep{castro_structure_1997,pereiro2001_a,lopez2003,ma2006,datta2007,perez_physical_2012} ground state where the respective other geometry is always very close in energy ($\sim$ 0.15\,eV) to the actual ground state. There is also no consensus on the magnetic ground state. We found reports of S\,=\,4 (see Ref.~[\onlinecite{castro_structure_1997}]), S\,=\,$\frac{11}{2}$ (see Ref.~[\onlinecite{guevara1997,andriotis1998,ma2006,sebetci_cobalt_2008}]), S\,=\,$\frac{13}{2}$ (see Ref.~[\onlinecite{fan_geometry_1997,pereiro2001_a,datta2007,ganguly2008,perez_physical_2012,kortus2002magnetic}]) and S\,=\,$\frac{15}{2}$ (see Ref.~[\onlinecite{lopez2003}]), where usually a S\,=\,$\frac{11}{2}$ or S\,=\,$\frac{13}{2}$ magnetic ground state is energetically very close to the actual magnetic ground state. There is also no connection between the ground state geometry and the magnetic ground state in literature. The experimental ionization potential about 6.2\,eV \citep{parks1990,yang1990} which is in good agreement with literature (IP\,$\sim$\,6\,eV) \citep{pereiro2001_a}, (IP\,=\,6.5\,eV) \citep{guevara1997} and the one observed in the present study (IP\,=\,6.1\,eV). Perez et al. \citep{perez_physical_2012} on the other hand report a significantly lower ionization potential of 5.1\,eV. The ionization potential (6.1\,eV) and the electron affinity (1.6\,eV) are nearly identical to the Co$_3$ and Co$_4$ clusters, hence we would expect a similar chemical reactivity. \\
As already noted for the tetramers, the pentamers of mixed composition show an increase in the total magnetic moment as well as a steady decrease in the binding energy with increasing Co content. For CoMo$_4$ we get a structure that is quite similar to the one of pure Mo$_5$ with a binding energy of 2.4\,eV. It could be described either as a trigonal bipyramid or as a square pyramid with a tilted base(see figure \ref{co3mo2}). We see again two very short Mo-Mo distances ($\sim$ 2.1\,$\mathring{\text{A}}$) indicating a dimerization of Mo as well as two longer bonds (2.68\,$\mathring{\text{A}}$) and two very long bonds (2.8\,$\mathring{\text{A}}$ and 2.9\,$\mathring{\text{A}}$). We found a S\,=\,$\frac{1}{2}$ magnetic ground state, a chemical potential and molecular hardness close to Mo$_5$ and Raman features at 29, 93, 117, 135, 171, 182, 251, 358 and 397\,cm$^{-1}$. For Co$_2$Mo$_3$ the ground state structure resembles a distorted square pyramid with a magnetic moment of S\,=\,2 and an easy plane anisotropy with a binding energy of 2.35\,eV. No dimerization of Mo is present, however the avarage Mo-Mo distance is smaller compared to the tetramers with more Mo content. Additionally we report Raman frequencies for this structure at 87, 104, 132, 155, 183, 232, 261, 296 and 382\,cm$^{-1}$. For Co$_3$Mo$_2$ the ground state structure has a binding energy of 2.36\,eV and resembles a rectangular pyramid as shown in figure \ref{co3mo2} with two Mo atoms at neighboring edges and two Co atoms occupying the remaining edges. The top atom is the residual Co atom. The Co-Co (2.20\,$\mathring{\text{A}}$) and Mo-Mo (2.17\,$\mathring{\text{A}}$) distance is very short and nearly identical. The Co-Mo distance in the plane is 2.32\,$\mathring{\text{A}}$ forming a nearly perfect
rectangular base of the pyramid. The dimerization of Mo is again present. The ground state has a magnetic moment of S\,=\,$\frac{5}{2}$ with an easy axis anisotropy of -1.6\,K, an ionization potential (6.2\,eV) and electron affinity (1.4\,eV) that resembles closely those of Co$_5$ and Raman features at 59, 105, 149, 204, 266, 299, 386\,cm$^{-1}$. For the Co$_4$Mo pentamer we found again a rectangular pyramid with a binding energy of 2.27\,eV and a magnetic moment of S\,=\,4. The anisotropy of D\,=\,-1.5 indicates an easy axis system with a barrier of 32\,K. The base of the pyramid is made up of Co atoms whereas the top atom is Mo as shown in figure \ref{co4mo}. The ionization potential (5.9\,eV) and the electron affinity (1.1\,eV) are very close to the respective values of the Co tetramer resulting in a nearly identical chemical reactivity. This indicates that the Mo atom in top is only loosely bound to the structure and does not strongly influence the chemical behavior of the cluster. We also report Raman features at 70, 121, 133, 166, 183, 247, 261, 281 and 349\,cm$^{-1}$.

\subsection{Hexamers}

\begin{figure}[h!]
  \centering
\subfloat [] {
 \includegraphics[scale=0.6]{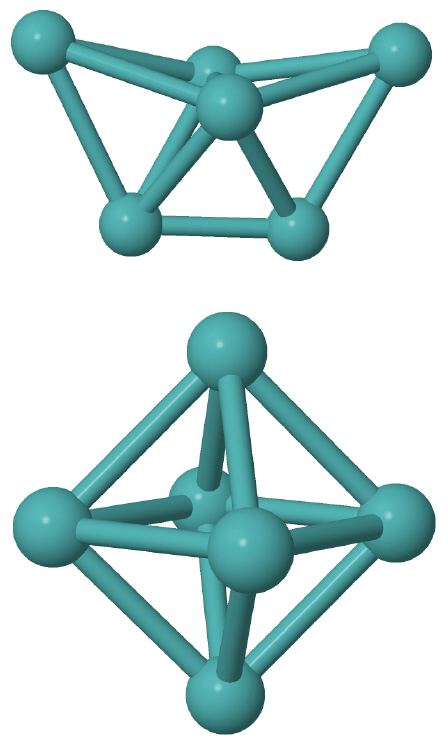}
\label{hexa}
}
 \subfloat [] {
 \includegraphics[scale=0.2]{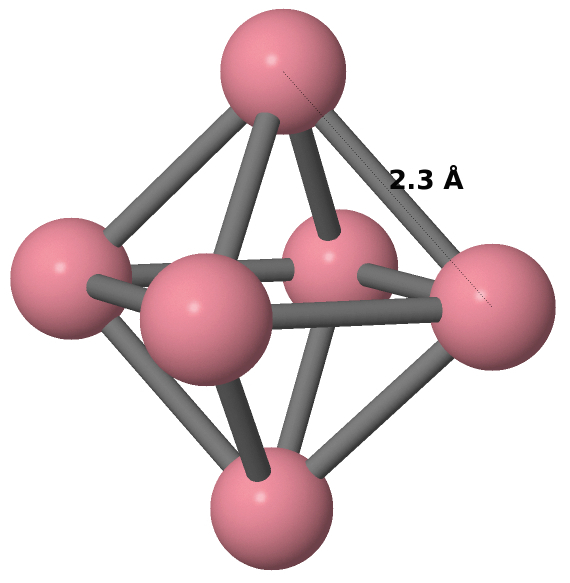}
\label{co6}
}
\\
 \subfloat [] {
 \includegraphics[scale=0.2]{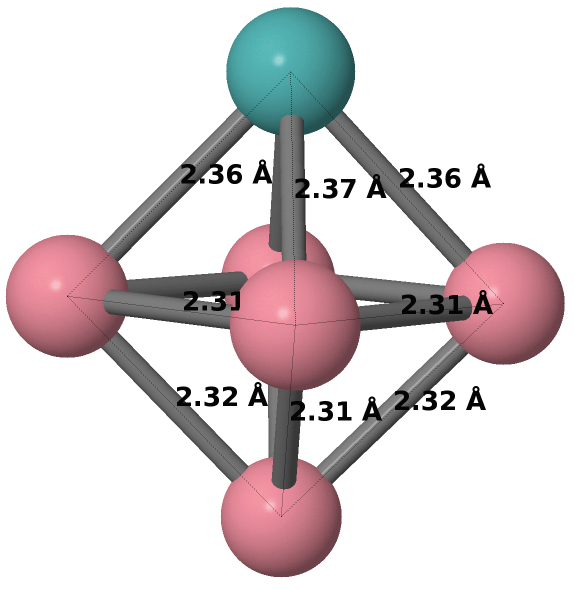}
\label{co5mo}
}
 \subfloat [] {
 \includegraphics[scale=0.2]{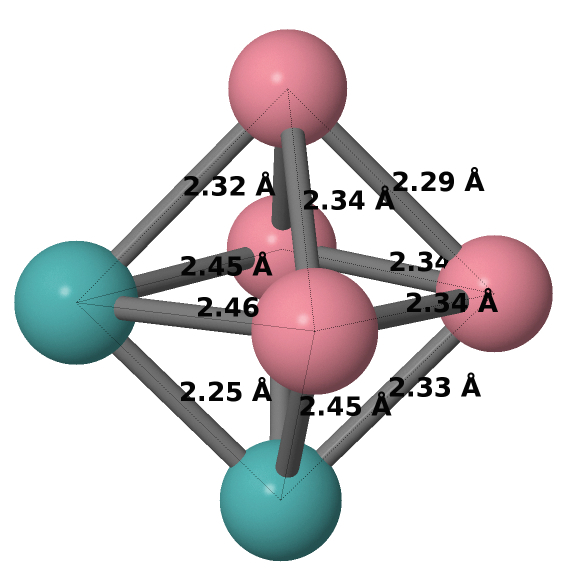}
 \label{co4mo2}
}
 \\
\subfloat [] {
 \includegraphics[scale=0.2]{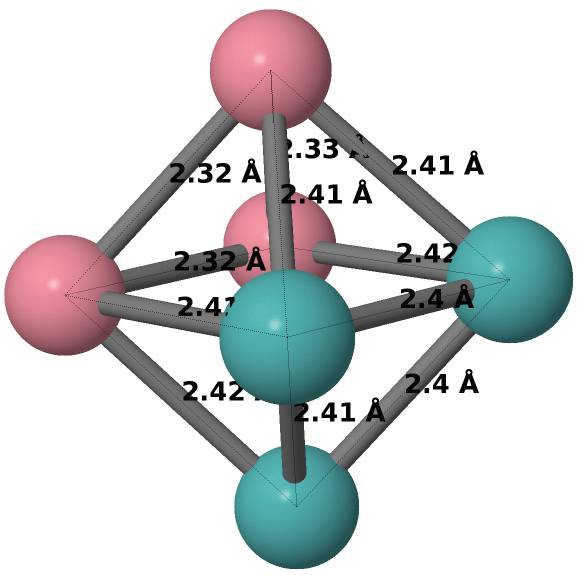}
 \label{co3mo3}
}
 \subfloat [] {
 \includegraphics[scale=0.2]{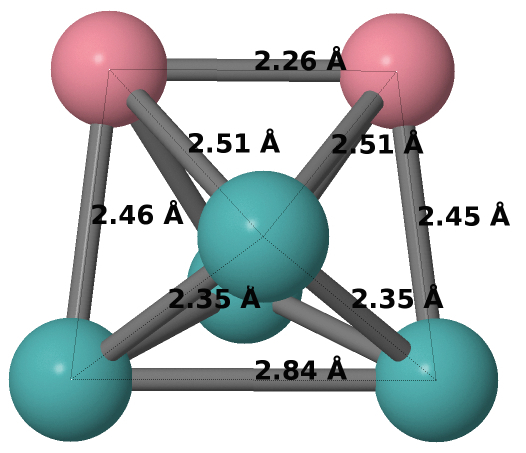}
 \label{co2mo4}
}
\\
 \subfloat [] {
 \includegraphics[scale=0.2]{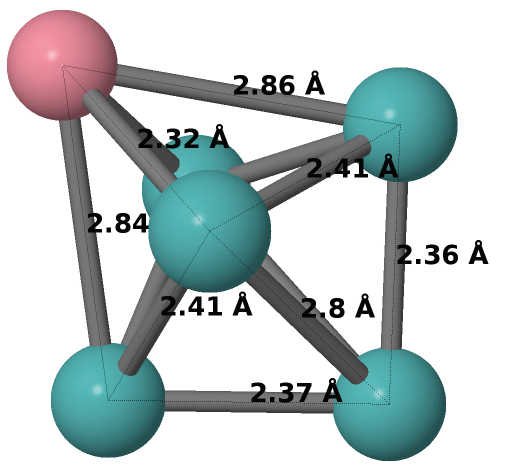}
 \label{como5}
}
 \subfloat [] {
 \includegraphics[scale=0.2]{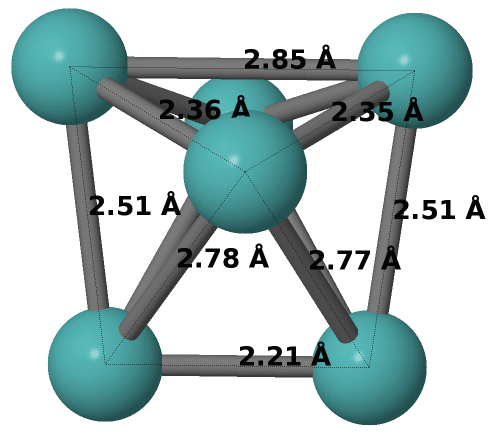}
 \label{mo6}
}
 \caption[]{a): Co$_6$, b): Co$_5$Mo, c): Co$_4$Mo$_2$, d): Co$_3$Mo$_3$, e): Co$_2$Mo$_4$, f): CoMo$_5$, g): Mo$_6$; all invisible bonds are as long as those directly in front of them. The
ground structure of Co$_6$ is a regular octahedron with an equidistant bond lenght of 2.3 $\mathring{\text{A}}$.  blue: Mo, red: Co}
 \label{hexamer}
 \end{figure}

As shown in figure \ref{hexa} we considered are two different possible starting geometries for the hexamer - an octahedral structure (12 possible bonds) and ship-like structure (11 possible bonds).
The respective ground state structure and selected properties of those structures are summarized in table \ref{overview}, whereas the ground state geometry can
be found in figure \ref{co6} - \ref{mo6}. 

 For the Mo$_6$ cluster we get a distorted ship-like structure as shown in figure \ref{mo6} which is already close to an octahedral structure with a binding energy of 2.8\,eV. The average Mo-Mo distance is 2.55\,$\mathring{\text{A}}$ including one very short distance of 2.21\,$\mathring{\text{A}}$ at the bottom of the ship-like structure. The structure is non magnetic with a very low electron affinity (0.1\,eV) and an ionization potential (5.2\,eV) comparable to the one of Mo$_5$. Accordingly we get a very low chemical potential ($\mu$\,=\,-2.67\,eV) in comparison to the other
structures. Furthermore we computed Raman frequencies at 62, 116, 140, 199, 206, 250, 341 and 365\,cm$^{-1}$. Min et al \citep{min2001} report an octahedral ground state structure with a buckled square base plane and a binding energy of 3.9\,eV, whereas other studies \citep{zhang2004,Koteski200579, gran2008} get a non magnetic deformed pentagonal pyramid with a binding energy of 2.92 - 3.16\,eV. \\
For the Co hexamer on the other hand we get a regular octahedral ground state with a length of 2.30\,$\mathring{\text{A}}$ for every bond and a binding energy of 2.73\,eV/atom. This regular octahedron has a S\,=\,7 magnetic ground state with and easy axis anisotropy of -0.02\,K. Many previous studies agree on the ground state to be a regular octahedron with S\,=\,7, a binding energy of 2.47 - 2.98\,eV/atom and a Co-Co distance of 2.27 - 2.37\,$\mathring{\text{A}}$ \citep{li_electronic-structure_1993,guevara1997,fan_geometry_1997,andriotis1998,datta2007,perez_physical_2012}. Only Andriotis et al report a significantly larger bond length of 2.67\,$\mathring{\text{A}}$ using tight-binding molecular dynamics. Other studies \citep{ma2006,sebetci_cobalt_2008} reported a distorted octahedron geometry with a S\,=\,7 magnetic ground state and an average Co-Co distance of 2.31 - 2.39\,$\mathring{\text{A}}$. Moreover we obtained a electron affinity (1.5\,eV) close to all previous pure Co$_n$ (n$>$2) clusters and an ionization potential of 6.6\,eV. Especially the IP is in good agreement with previous experimental (IP\,$\sim$\,6.2\,eV) \citep{yang1990,parks1990} and theoretical (IP\,=\,5.07\,eV \citep{perez_physical_2012} and IP\,=\,6.9\,eV \citep{guevara1997}) work. Moreover we observed Raman features at 193, 238 and 347\,cm$^{-1}$. Sebetci et al. also computed optical frequencies (94 and 296\,cm$^{-1}$), however they predict a distorted octahedron for the ground state, hence the results may not be comparable.\\
For the mixed hex-atomic clusters we observe a nice transition from the distorted ship-like structure of pure Mo$_6$ to a regular octahedron ground state structure as observed for Co$_6$ with
increasing Co-content (see figure \ref{hexamer}). This is accompanied by an increase in the total magnetic moment as well as a steady decrease in the binding energy with increasing Co content in
agreement with the trends already reported here for smaller mixed clusters. For CoMo$_5$ we also get a distorted ship-like ground state structure with S\,=\,$\frac{1}{2}$ and a binding energy of 2.74\,eV/atom. As shown in figure \ref{como5} the upper left edge is occupied by the Co atom which has two short bonds (2.32\,$\mathring{\text{A}}$) to the middle Mo atoms and two long bonds ($\sim$\,2.85\,$\mathring{\text{A}}$) to the neighboring edges. It could also be viewed as a strongly distorted octehedron with a buckled square base plane where the buckling is due to the Co atoms that drags two Mo atoms out of the plane. There are no very short Mo-Mo distances hence no dimerization occurs. Instead we see six medium size distances ($\sim$\,2.38\,$\mathring{\text{A}}$) and two very long bonds of 2.8\,$\mathring{\text{A}}$. Furthermore we found an ionization potential of 5.7\,eV and a electron affinity of 1.1\,eV as well as pronounced Raman features at 69, 142, 200, 218, 284 and 351\,cm$^{-1}$.
For Co$_2$Mo$_4$ the ground state structure can also be described as a distorted ship-like arrangement with an S\,=\,2 ground state and an easy axis anisotropy of -10.8\,K. Here the upper two edges of the
ship are occupied by Co atoms forming a short Co-Co bond of length 2.26\,$\mathring{\text{A}}$ (see figure \ref{co2mo4}). They form two shorter bonds (2.45\,$\mathring{\text{A}}$) to the remaining two edges and four longer, equidistant bonds (2.51\,$\mathring{\text{A}}$) to the center Mo atoms thereby pushing them downwards. This results in four short Mo-Mo bonds (2.35\,$\mathring{\text{A}}$) and one very long Mo-Mo bond (2.84\,$\mathring{\text{A}}$). Again this structure could be also described as a distorted octehedron with a buckled square base plane, where the buckling is due to the Co atoms as already discussed for the CoMo$_5$ hexamer. The ionization potential of 5.6\,eV and electron affinity (0.5\,eV) are lower compared to the CoMo$_5$ cluster and we report Raman frequencies at 94, 115, 148, 177, 194, 225, 184 and 356\,cm$^{-1}$. The Co$_3$Mo$_3$ cluster on the other hand resembles clearly a distorted octahedron with a binding energy of 2.72 ev/atom. Again there is no dimerization of the Mo atoms observed, on the contrary all Mo-Mo bonds are equidistant (2.4\,$\mathring{\text{A}}$). The same applies for the Co-Co bonds (2.32\,$\mathring{\text{A}}$) although those are shorter than the Mo-Mo bonds. The Co-Mo bonds are also all equidistant at 2.42\,$\mathring{\text{A}}$. The ionization potential of 5.7\,eV is very close to the one of Co$_2$Mo$_4$, whereas the electron affinity is considerably larger (1.6\,eV). Furthermore we see Raman features at 131, 172, 230, 293 and 364\,cm$^{-1}$, a magnetic ground state of S\,=\,$\frac{5}{2}$ and an easy plane anisotropy (D\,=\,3.6\,K). The Co$_4$Mo$_2$ ground state structure is also clearly an octahedron with a binding energy of 2.66\,eV/atom, a magnetic moment of S\,=\,3 and an easy axis anisotropy (D\,=\,-2.7\,K). We report a very short Mo-Mo distance (2.25\,$\mathring{\text{A}}$) indicating the already discussed dimerization of Mo due to its half-filled shell. The average Co-Co distance is 2.33\,$\mathring{\text{A}}$ with only slight deviations for the different Co-Co distances (2.29 - 2.34\,$\mathring{\text{A}}$). The Co-Mo distance on the other hand show a greater variation (2*2.33\,$\mathring{\text{A}}$ and 4*2.45\,$\mathring{\text{A}}$) averaging to 2.41\,$\mathring{\text{A}}$. The computed Raman frequencies are 110, 135, 144, 161, 187, 208, 219, 252, 298 and 365\,cm$^{-1}$. The chemical reactivity ($\mu, \eta$)
is nearly identical to the one of Co$_3$Mo$_3$ due to the same calculated electron affinity and ionization potential. Co$_5$Mo finally resembles almost a regular octahedron. The Mo atom sits at the
top of the octahedron and is 2.37\,$\mathring{\text{A}}$ away from the neighboring Co atoms. All the Co-atoms are also nearly equidistant with 2.31\,$\mathring{\text{A}}$ and 2.32\,$\mathring{\text{A}}$ bond length. We found a S\,=\,$\frac{9}{2}$ magnetic ground state with an easy plane anisotropy of -0.6\,K and a binding energy of 2.57\,eV/atom which is the lowest observed for the hexamers. Again the chemical reactivity is closely related Co$_4$Mo$_2$ and Co$_3$Mo$_3$ due to the nearly identical ionization potential (5.9\,eV) and electron affinity (1.4\,eV). Additionally we predict Raman features at 146, 158, 216 and 338\,cm$^{-1}$. 

\section{General Remarks}

As already indicated within the discussion of the various-sized clusters there are some general trends that can be observed.
 \begin{figure}
     \begin{center}
      \includegraphics[height=0.3\textwidth ,keepaspectratio=true]{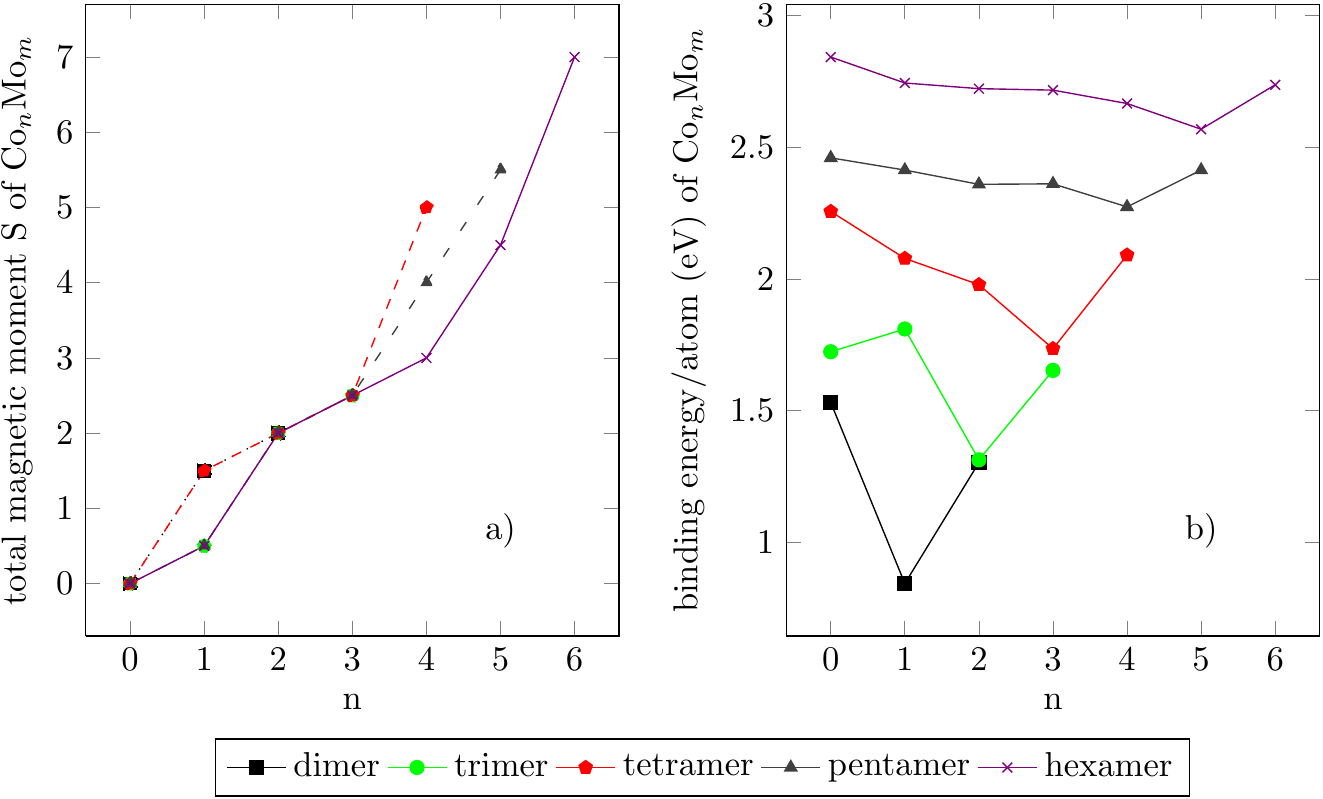}
      \caption{Various cluster properties with respect to the Mo-concentration: a) binding energy (eV/atom), b) total magnetic moment S, c) electron affinity (eV), d)
chemical potential $\mu$ (eV), e) molecular hardness $\eta$ (EV) and f) HOMO-LUMO gap (eV)}
     \end{center}
\end{figure}
As shown in figure 5a) all pure Mo complexes show non magnetic behavior. All pure Co complexes show a magnetic ground state that can be described by 
\begin{equation}
S = \begin{cases} \frac{2n+1}{2} &\mbox{if } n = \text{odd} \\ 
\frac{2n+2}{2} & \mbox{if } n = \text{even} \end{cases}  \label{magnetization}
\end{equation}
where n is the amount of Co in a given cluster \citep{sebetci_cobalt_2008}. Consequently the largest magnetic ground state (S\,=\,7) is observed for the Co$_6$ cluster. For the mixed clusters we observe
an
increase in the magnetic moment with increasing Co content, where the magnetic moment is in general due the Co atoms. The magnetic ground state of these clusters can in general also be predicted by equation (\ref{magnetization}). However in some cases the magnetic moment is quenched, hence a smaller magnetic ground state is found. This seems to be the case for all mixed clusters with a Co content
of n $\ge$ 4 and also for n\,=\,1, where we get an S\,=\,$\frac{1}{2}$ ground state for CoMo$_2$,CoMo$_4$ and CoMo$_5$ instead the expected S\,=\,$\frac{3}{2}$ ground state predicted by the model. Another
interesting trend is the steady decrease of the strength of the magnetic anisotropy
(regardless of a possible change in the sign of D) with increasing magnetic moment. The only exception are the tetramers where no clear trend is visible. \\
As shown in figure 5b), there is also a steady increase of the binding energy per atom with increasing cluster size, where the pure Mo clusters are usually a bit more stable than the pure Co
clusters for a given total number of atoms x in a cluster. This trend is to be expected as the coordination number increases with increasing cluster size. Starting from pure Mo clusters the binding
energy goes down with increasing Co content (see figure \ref{fig:dist}) . For each cluster size there is a slight decrease in the binding energy with increasing Co content as already discussed in the previous sections, where the
mixed cluster with only one Mo atom is always the least stable one. \\
The change in the ionization potential and electron affinity with increasing cluster size can be discussed within the model of conducting spherical droplets as explained elsewhere
\citep{smith1965,wood1981,Schumacher1984,morse1986}. Within this model the cluster is treated as a conducting sphere of radius R and the change of the electron affinity (EA) and the ionization
potential (IP) is given by: 
\begin {eqnarray}
 \text{IP}(\text{R})&\,=\,&\text{W}+\frac{3}{8} \frac{\text{e}^2}{\text{R}}\\
 \text{EA}(\text{R})&\,=\,&\text{W}-\frac{5}{8} \frac{\text{e}^2}{\text{R}},
\end {eqnarray}
with respect to the charge e and the bulk work function W, which contains the intrinsic information regarding the Fermi level of the infinite bulk material \citep{knickelbein1999}. For very large
spheres IP as well as EA should approach this value. This relation can be rewritten in terms of the volume V for a spherical x-atomic cluster \citep{morse1986}: 
\begin {eqnarray}
 \text{IP}(\text{x})&\,=\,&\text{W}+8.7\text{x}^{-\frac{1}{3}}\text{V}^{-\frac{1}{3}}\\
 \text{EA}(\text{x})&\,=\,&\text{W}-14.5\text{x}^{-\frac{1}{3}}\text{V}^{-\frac{1}{3}}.
\end {eqnarray}
The main conclusion of this model is that one would expect an increase of the electron affinity and an decrease of the ionization potential with increasing cluster size. As shown in figure table
\ref{overview}, this is in general true for the electron affinity and applies also for the ionization potential. It is also very
interesting,
that for a given cluster size x the ionization potential and the electron affinity increases with increasing Co content. As the chemical potential $\mu$ and the molecular hardness $\eta$ are
computed using the ionization potential and the electron affinity (see equation \ref{potential} and \ref{hardness}), the trends can also be discussed with respect to the
conducting sphere droplet model. Within this model we get: 
\begin {eqnarray}
 \mu(\text{x})&\,=\,&-\frac{1}{2}(2\text{W}-5.8\text{x}^{-\frac{1}{3}}\text{V}^{-\frac{1}{3}})\\
 \eta(\text{x})&\,=\,&\frac{1}{2}(23.2\text{x}^{-\frac{1}{3}}\text{V}^{-\frac{1}{3}}),
\end {eqnarray}
where a decrease of the chemical potential $\mu$ is predicted with increasing cluster size x. This works as a rough estimate to describe the evolution of $\mu$ with increasing
cluster size x. Another interesting trend is the increase of the chemical potential with increasing Co-content for a given cluster size x. This relates directly to the increase in the electron
affinity with increasing Co-content as discussed before and shown in figure 5c). Accordingly the cluster become more stable with increasing Co-content. The molecular hardness $\eta$ is also predicted
to decrease with increasing cluster size x due to the x$^{-\frac{1}{3}}$ dependency. This works very well for small Co-contents in the clusters and becomes more diffuse for n$\ge$ 2. The reduced
molecular hardness for larger clusters results directly in a larger polarizability for these clusters. \\
The behavior of the HOMO-LUMO gap is also very interesting. The gap for pure Mo clusters is considerably larger than the gap for pure Co clusters of the same size.
Additionally there is an even-odd effect traceable for all the pure Mo cluster which is also reported in literature \citep{zhang2004}. The same effect, also to a lesser extent is visible for the pure
Co clusters. For all clusters of size x (Co$_\text{n}$Mo$_\text{m}$ clusters, x\,=\,m+n) we observe a steady decrease of the gap with increasing Co content. The exception is x\,=\,5 where again even-odd
jumps are visible and x\,=\,6, where the gap of the pure Co hexamer is larger than the one of Co$_5$Mo. On the other hand we see for n $\ge$ 2 in the Co$_\text{n}$Mo$_\text{m}$ cluster an increase
in the HOMO-LUMO gap with increasing cluster size. \\
As discussed in great detail by Baletto et al \citep{Baletto2005} the average bond length is supposed to shrink with decreasing size of the cluster. This is true for all the discussed clusters and
most evident at the evolution of the average Mo-Mo distance for pure Mo clusters (see figure 6c).
 \begin{figure}
     \begin{center}
      \includegraphics[height=0.4\textwidth ,keepaspectratio=true]{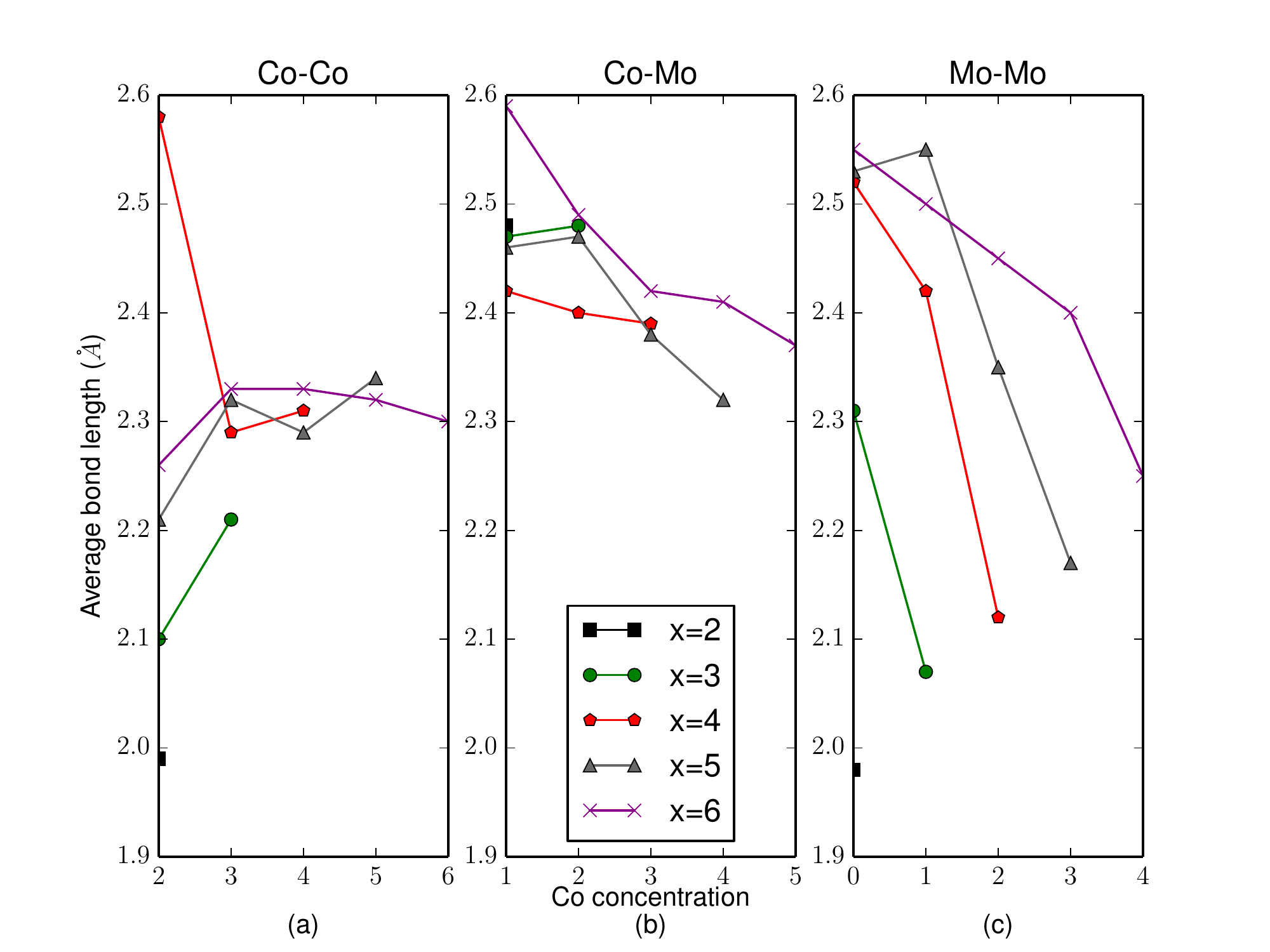}
      \caption{Average bond distances between neighboring (a) Co atoms, (c) Mo atoms and (b) Co and Mo atoms.}
      \label{fig:dist}
     \end{center}
\end{figure}
For a given cluster size x we also observe a decrease in the average Mo-Mo and Co-Mo distance with increasing Co content. The results on the Co-Co distance are not so clear. For one we see a steady
increase in the Co-Co distance for pure Co complexes with the average Co-Co distance of the hexamer slightly lower then the average distance in the pentamer and tetramer, which is in agreement with
previous studies \citep{datta2007}. There are also no clear trends, however for a fixed n we usually see an increase in the average Co-Co distance with increasing cluster size x. 

\section{Conclusions}
Overall it can be stated that the usual, size-dependent tendencies for metallic clusters are fullfilled. We observe an increase in the binding energy, electron affinity, and average bond length with
increasing cluster size as well as a decrease in the ionization potential, chemical potential, molecular hardness and the HOMO-LUMO gap. The evolution of the electronic properties (electron affinity,
ionization potential) and chemical reactivity (chemical potential, molecular hardness) with increasing cluster size x can be well understood within the conducting sphere model. The magnetic ground
state is mainly governed by the Co atoms in a given Co$_\text{n}$Mo$_\text{m}$ cluster and increases with increasing cluster size and increasing Co content in a cluster of size x. For nearly all
systems we observe a decrease in the magnetic anisotropy (regardless of a possible change in the sign of D) with increasing magnetic moment, the only exception being x\,=\,4 where no clear trend is
visible. This has also been shown experimentally for small Co-clusters \citep{gambardella_giant_2003}. The trends observed for clusters of size x with different chemical compositions can be assigned
to
the Co content n and for a given n clusters of different sizes usually behave likewise. For example the binding energy for n\,=\,0 increases with increasing cluster size. This rule applies also for
n\,=\,1,..,6. Similar conclusion can be made for the other properties considered in the present paper like electron affinity, chemical potential, bond length and so on. Due to the very interesting
magnetic properties these Co$_\text{n}$Mo$_\text{m}$ complexes might also be interesting in the field molecular electronics, for example for data storage or spin-dependent transport. 

\begin{acknowledgments}
The authors want to thank the National Science Foundation (XSEDE) for the computational resources, the NordForsk Network on Nanospintronics and the Department of Energy (BES), the Department of Defence (ONR-Global \& ITC-ATL) for the financial support of the Summer School "The Como Moments: Theoretical \& Computational Modeling of Magnetically Ordered Molecules \& Electronic Nano-Transport of Spins: State of Art and Unanswered Questions" within which the idea of this paper was developed. The authors also want thank the ZIH Dresden and the Cluster of Excellence "Structure Design of Novel High-Performance Materials via Atomic Design and Defect Engineering (ADDE)" which is financially supported by the European Union (European regional development fund) and by the Ministry of Science and Art of Saxony (SMWK) for the computational resources to do the calculations presented in this work. 

\end{acknowledgments}

% \printbibliography[heading=bibintoc]
%\newpage
\bibliography{comopaper.bib}
%\bibliographystyle{unsrt}
% \begin{thebibliography}{99}
% 
% \end{thebibliography}
\end{document}